\documentclass[prd,twocolumn,nofootinbib,preprintnumbers]{revtex4}
\usepackage{color}
\usepackage{latexsym}
\usepackage{revsymb}
\usepackage{amsmath,amssymb}
\usepackage{natbib}
\usepackage{feynmp}
\def\Slash#1{\hskip 0.05 cm \slash\hskip -0.2 cm #1} 
\def\SSlash#1{\hskip 0.1 cm \slash\hskip -0.3 cm #1}

\begin{document}
\title{Rate of gravitational inflaton decay via gauge trace anomaly}
\author{Yuki Watanabe}
\affiliation{Arnold Sommerfeld Center for Theoretical Physics, Ludwig-Maximilians-University, Theresienstrasse 37, 80333 Munich, Germany}
\email{yuki.watanabe@physik.lmu.de}
\date{\today}
\preprint{LMU-ASC 72/10}

\begin{abstract}
  We analyze decay processes of the inflaton field, $\phi$, during the coherent oscillation phase after inflation in $f(\phi)R$ gravity.
It is inevitable that the inflaton decays gravitationally into gauge fields in the presence of $f(\phi)R$ coupling.
We show a concrete calculation of the rate that the inflaton field decays into a pair of gauge fields via the trace anomaly.
Comparing this new decay channel via the anomaly with the channels from the tree-level analysis, we find that the branching ratio crucially depends on masses and the internal multiplicities (flavor quantum number) of decay product particles.
While the inflaton decays exclusively into light fields, heavy fields still play a role in quantum loops.
We argue that this process in principle allows us to constrain the effects of arbitrary heavy particles in the reheating.
We also apply our analysis to Higgs inflation, and find that the gravitational decay rate would never exceed gauge interaction decay rates if quantum gravity is unimportant.  
\end{abstract}
\maketitle

\section{Introduction}\label{sec:introduction}

Inflationary cosmology has passed a number of stringent observational tests, such as observations of cosmic microwave background temperature anisotropy \cite{Komatsu:2010fb}. Despite the success of inflation models, the identity of the field that drives inflation is little understood.  Any inflation model requires a graceful exit, the so-called reheating after inflation (see reviews, e.g., \cite{Linde:1990,Lyth:1998xn,Mukhanov:2005sc,Weinberg:2008zzc,Braden:2010wd} and references therein); otherwise the observed Universe cannot be predicted by the models. 
The Universe must be thermalized and dominated by radiation fluids before the primordial nucleosynthesis. 
Thermalization is initiated by the decay of the inflaton field into lighter particles, and then the particles come to a state of local equilibrium with each other.

It has been shown that inflation occurs naturally in models with nonminimal gravity \cite{Accetta:1985du, La:1989za,Futamase:1987ua,Salopek:1988qh,Fakir:1990eg}, and the spectrum of scalar curvature perturbations \cite{Makino:1991sg,Kaiser:1994vs,Faraoni:2000nt} as well as of tensor gravity wave perturbations \cite{Komatsu:1997hv,Komatsu:1999mt,Hwang:1998fi,Tsujikawa:2004my} can be affected by the presence of $f(\phi)R$, thereby allowing us to constrain $f(\phi)$ from the cosmological data.
The $f(\phi)R$ gravity can also play an important role in the inflaton decay at the reheating \cite{Watanabe:2006ku,Watanabe:2007tf,Kaloper:2008gs,Pallis:2010wt}.

In theories with nonminimal couplings between the Ricci curvature and scalar fields, e.g., scalar-tensor gravity \cite{Watanabe:2006ku,Watanabe:2007tf,Kaloper:2008gs,Pallis:2010wt}, $\mathcal{R}^2$ gravity \cite{Starobinsky:1980te,Vilenkin:1985md,Mijic:1986iv,Kalara:1990ar,Faulkner:2006ub,Gorbunov:2010bn}, supergravity \cite{Endo:2006qk,Endo:2007sz,Endo:2007ih}, and higher dimensional gravity theories, inflaton fields can decay via gravitational effects.
The gravitational two-body decay rate is typically given by
\begin{eqnarray}
\Gamma_{\rm grav}(\sigma\to 2) \approx \frac{C}{16\pi}\frac{m_{\sigma}^3}{M_{Pl}^2},
\end{eqnarray}
where $M_{Pl}\equiv (8\pi G)^{-1/2}\simeq 2.4 \times 10^{18}$ GeV, $m_{\sigma}$ is the inflaton mass, and $C$ is a model-dependent fudge factor.
The inflaton field condensate, $\sigma$, decays into any pair of light fields if it is not conformally invariant.
The unknown factor, $C$, significantly depends not only on the gravitational sector but also on the detailed properties of the matter sector, such as mass spectrum, spins, and  the number of degrees of freedom at the energy scale of reheating. 
Thus, understanding physical grounds of $C$ merits further understanding of the reheating.

In this paper, we will take advantage of this gravitational nature in the reheating mechanism. 
We consider only the gravitational effect and assume no direct interaction between the inflaton field and matter fields. 
Since gravity interacts universally, the inflaton field interacts with every field that exists at the reheating.
Especially, we focus on an emergent interaction in the case where the inflaton couples to gravity nonminimally. 
Although the inflaton is assumed to be gauge-singlet, its decay products (matter and radiation) may be charged under some gauge group, e.g., $SU(N)$.
We aim to constrain physical properties of new particle species at the reheating, such as spin, mass spectrum, and the number of degrees of freedom. 
As was discussed in \cite{Watanabe:2007tf}, the large number of degrees of freedom significantly enhances the gravitational decay.

Our approach can be contrasted with the recent development of the Higgs inflation model \cite{Bezrukov:2007ep,DeSimone:2008ei,Bezrukov:2009db,Barvinsky:2009ii,Bezrukov:2010jz}, where no new degrees of freedom are introduced. The model with nonminimal kinetic gravitational coupling, $G^{\mu\nu}\partial_{\mu}\phi\partial_{\nu}\phi$, is another possibility \cite{Germani:2010gm,Germani:2010ux} in this direction.
Reheating in Higgs inflation has been considered by \cite{GarciaBellido:2008ab,Bezrukov:2008ut}.

The organization of this paper is as follows.
In Sec.~\ref{sec:induced_coupling} an inflation model with nonminimal gravitational coupling and $U(1)$-charged matter Lagrangian is given. The gravitationally induced interactions are derived.
The rates of gravitational inflaton decay are given in Sec.~\ref{sec:inflaton_decay}.
Our main result, the decay rate via conformal (trace) anomaly [Eq.~(\ref{eq:fs-loop})], is presented in subsection~\ref{subsec:anomaly_decay}.
Implications of the gravitational inflaton decay for reheating are investigated in Sec.~\ref{sec:implication}.
An implication for Higgs inflation is given there. 
We show a detailed derivation of fermionic one-loop decay rate via trace anomaly in the Appendix.
We work with the metric signature $(+,-,-,-)$ throughout the paper. This sign convention is different from that in previous works \cite{Watanabe:2006ku,Watanabe:2007tf}.

\section{$f(\phi)R$ gravity and induced couplings}\label{sec:induced_coupling}

For concreteness, we assume that the inflaton, $\phi$, is a neutral scalar field that is nonminimally coupled to gravity but gauge-singlet, i.e., no direct interaction with the $SU(3)_C\times SU(2)_L\times U(1)_Y$ standard model sector or any other matter fields. To make physics clear, we employ a QED-like model, $U(1)$-charged scalars and fermions, as a matter sector at the energy scale of reheating:
\begin{eqnarray}
\mathcal{L}&=&\sqrt{-g}\left[-\frac12 f(\phi)R
+\frac12 g^{\mu\nu}\partial_{\mu}\phi\partial_{\nu}\phi-V(\phi)\right] 
+\mathcal{L}_{\rm m},\nonumber\\
\mathcal{L}_{\rm m}&=&\sqrt{-g}\left[
\sum_{s=1}^{N_{\chi}}\left((D_{\mu}\chi_s)^*D^{\mu}\chi_s - m^2_{s}\chi_s^*\chi_s\right)\right. \nonumber\\
&& \left. +\sum_{f=1}^{N_{\psi}}\bar{\psi}_f\left(i\SSlash{D} - m_f\right)\psi_f
-\frac14 F_{\mu\nu}F^{\mu\nu}\right],
\end{eqnarray}
where $N_{\chi}$ and $N_{\psi}$ are the internal flavor quantum number for scalar and fermion, respectively.
We impose the boundary condition $f(v)=M_{Pl}^2$, where $v$ is the vacuum expectation value (vev) of $\phi$ at the potential minimum, to guarantee the ordinary Einstein gravity at low energy.\footnote{This dynamic boundary condition is motivated by the spirit of Zee's induced gravity theory \cite{zee:1978wi} and also applied to other models of nonminimally coupled inflation \cite{Accetta:1985du,Kaloper:2008gs}.}
We have explicitly shown the square root of the determinant of the metric, $\sqrt{-g}$, to emphasize that two sectors talk to each other only minimally via gravity. 
The covariant derivatives for each field are defined as
\begin{eqnarray}
D^{\mu}\chi &\equiv& g^{\mu\nu}(\partial_{\nu}-igA_{\nu})\chi,\nonumber\\
\SSlash{D}\psi &\equiv& e^{\mu}{}_{\alpha}\gamma^{\alpha}(\partial_{\mu}-\Gamma_{\mu}-igA_{\mu})\psi,\nonumber\\
F^{\mu\nu} &\equiv& g^{\mu\rho}g^{\nu\sigma}F_{\rho\sigma} 
\equiv g^{\mu\rho}g^{\nu\sigma}(\nabla_{\rho}A_{\sigma}-\nabla_{\sigma}A_{\rho})\nonumber\\
&=& g^{\mu\rho}g^{\nu\sigma} (\partial_{\rho}A_{\sigma}-\partial_{\sigma}A_{\rho}),
\end{eqnarray}
where $g$ is the gauge coupling constant and assumed to be weakly coupled, $g \lesssim \mathcal{O}(1)$. Here $\alpha,\,\beta,\,\gamma,\,\cdots$ denote Lorentz indices while $\lambda,\,\mu,\,\nu,\,\cdots$ denote general coordinate indices. $e^{\mu}{}_{\alpha}$ is a tetrad (vierbein) field. $\Gamma_{\mu}$ is a spin connection and $\Sigma^{\alpha\beta}$are generators of the Lorentz group given by $\Gamma_{\mu}\equiv -\frac12\Sigma^{\alpha\beta}e^{\lambda}{}_{\alpha}\nabla_{\mu}e_{\lambda\beta}$ and $\Sigma^{\alpha\beta}=-\Sigma^{\beta\alpha}=\frac14[\gamma^{\alpha},\gamma^{\beta}]$ (see, e.g., Sec.~12.5 in \cite{Weinberg:1972gc}; Sec.~3.8 in  \cite{Birrell:1982ix}).
This simple model is rich enough to demonstrate important gravitational decay processes in reheating as we will see in the following section.

After inflation the inflaton field oscillates about the minimum of the potential. Thus we expand $\phi$ as 
\begin{eqnarray}
\phi = v+\sigma, 
\end{eqnarray}
where $\sigma$ represents coherent condensate of inflaton quanta measured from its vev, $\langle\phi\rangle=v$.
In the Jordan frame, where the theory was originally defined, the inflaton, $\sigma$, decays into the matter sector through loops and mixing involving graviton if there is no direct interaction. 
It is simpler to compute physical quantities in the Einstein frame, where inflaton and graviton are diagonalized.

Performing conformal transformation to the metric tensor 
\begin{eqnarray}
g_{\mu\nu}&\to& \hat{g}_{\mu\nu}=\Omega^2 g_{\mu\nu},\quad
\Omega^2 = \frac{f(\phi)}{M_{Pl}^2},
\end{eqnarray}
and rescaling fields simultaneously,\footnote{Since canonical kinetic terms are needed to quantize fields in the new coordinate (Einstein frame), the field redefinitions~(\ref{eq:rescale}) are necessary. The rescaling of $\sigma$ follows immediately from \cite{Maeda:1988ab}
\begin{eqnarray}
\phi \to \hat{\phi}= M_{Pl}\int^{\phi}d\phi \sqrt{\frac{1}{f(\phi)}+\frac32\left(\frac{f'(\phi)}{f(\phi)}\right)^2}.\nonumber
\end{eqnarray}
}
\footnote{In the linearized theory, infinitesimal conformal transformation simply corresponds to diagonalization between graviton and inflaton, $\hat{\sigma}$, canonically normalized. Since two frames are very close at the potential minimum of the inflaton, we do not have to rescale matter fields \cite{Watanabe:2006ku}. If one would like to know nonlinear interaction, finite conformal transformation is necessary and matter fields should also be rescaled accordingly.}
\begin{eqnarray}
\sigma &\to& \hat{\sigma}= \sigma\sqrt{1+\frac32\left(\frac{f'(v)}{M_{Pl}}\right)^2},\quad
\chi \to \hat{\chi}=\Omega^{-1}\chi,\quad \nonumber\\
\psi &\to& \hat{\psi}=\Omega^{-3/2}\psi,\nonumber\\
A_{\mu}&\to& \hat{A}_{\mu}=A_{\mu},\quad A^{\mu}\to \hat{A}^{\mu}=\Omega^{-2}A^{\mu},
\label{eq:rescale}
\end{eqnarray} 
interactions are spontaneously induced by gravity. 
Transformed matter Lagrangians are given by
\begin{eqnarray}
\hat{\mathcal{L}}_{\chi}= \sqrt{-\hat{g}}\sum_{s=1}^{N_{\chi}}\left[\hat{g}^{\mu\nu}(\mathcal{D}_{\mu}\hat{\chi}_s)^*\mathcal{D}_{\nu}\hat{\chi}_s - \Omega^{-2}m_{s}^2\hat{\chi}^*_s\hat{\chi}_s\right],\label{eq:lag_scalar}\\
\hat{\mathcal{L}}_{\psi}= \sqrt{-\hat{g}}\sum_{f=1}^{N_{\psi}}\hat{\bar{\psi}}_f\left[i\hat{e}^{\mu}{}_{\alpha}\gamma^{\alpha}(\partial_{\mu}-\hat{\Gamma}_{\mu}-ig\hat{A}_{\mu})\right.\nonumber\\
 \left. -\Omega^{-1}m_f\right]\hat{\psi}_f,\label{eq:lag_fermion}\\
\hat{\mathcal{L}}_{A_{\mu}}=  -\frac14\sqrt{-\hat{g}} \hat{g}^{\mu\rho}\hat{g}^{\nu\sigma}\hat{F}_{\mu\nu}\hat{F}_{\rho\sigma},\hspace{2cm}\label{eq:lag_gauge}
\end{eqnarray}
where the covariant derivative for scalars is defined as
\begin{eqnarray}
\mathcal{D}_{\mu}\hat{\chi} 
\equiv \partial_{\mu}\hat{\chi}+\hat{\chi}\partial_{\mu}(\ln \Omega)-ig\hat{A}_{\mu}\hat{\chi}.
\end{eqnarray}
From Eqs.~(\ref{eq:lag_scalar}), (\ref{eq:lag_fermion}) and (\ref{eq:lag_gauge}) it can be seen that massless fermions and gauge bosons are conformally invariant at the classical level while massless scalars are not.\footnote{
The spinor and gauge field connections are conformally invariant:
\begin{eqnarray}
\hat{\Gamma}_{\mu} 
&\equiv&  -\frac12\Sigma^{\alpha\beta}\hat{e}^{\lambda}{}_{\alpha}\hat{\nabla}_{\mu}(\hat{e}_{\lambda\beta})\nonumber\\
&=& -\frac12\Sigma^{\alpha\beta}\Omega^{-1}e^{\lambda}{}_{\alpha}(\partial_{\mu}e_{\lambda\beta}-\hat{\Gamma}^{\sigma}{}_{\mu\lambda}e_{\sigma\beta})\nonumber\\
&&-\frac12\Sigma^{\alpha\beta}e^{\lambda}{}_{\alpha}e_{\lambda\beta}\Omega^{-1}\partial_{\mu}\Omega \nonumber\\
&=& \Gamma_{\mu},\nonumber\\
\hat{\Gamma}^{\sigma}{}_{\mu\lambda} 
&\equiv& \frac12 \hat{g}^{\nu\sigma}(\partial_{\lambda}\hat{g}_{\mu\nu}+\partial_{\mu}\hat{g}_{\lambda\nu}-\partial_{\nu}\hat{g}_{\mu\lambda})\nonumber\\
&&\hspace{-1cm}= \Gamma^{\sigma}{}_{\mu\lambda} + (\delta^{\sigma}{}_{\mu}\Omega^{-1}\partial_{\lambda}\Omega + \delta^{\sigma}{}_{\lambda}\Omega^{-1}\partial_{\mu}\Omega-g^{\nu\sigma}g_{\mu\lambda}\Omega^{-1}\partial_{\nu}\Omega).\nonumber\\
\hat{F}_{\mu\nu} &=&  {F}_{\mu\nu},\quad 
\hat{F}^{\mu\nu} = \Omega^{-4}{F}^{\mu\nu}.\nonumber
\end{eqnarray}
}

Interaction between the inflaton field and matter fields appears through the conformal factor $\Omega(\hat{\sigma})$,
\begin{eqnarray}
\Omega^2 &=& 1+\frac{f'(v)\sigma}{M_{Pl}^2}+\frac{f''(v)\sigma^2}{2M_{Pl}^2}+\frac{f'''(v)\sigma^3}{6M_{Pl}^2}+\cdots \nonumber\\
&=& 1+\frac{F_1(v)\hat{\sigma}}{M_{Pl}^2}+\frac{F_2(v)\hat{\sigma}^2}{2M_{Pl}^2}+\frac{F_3(v)\hat{\sigma}^3}{6M_{Pl}^2}+\cdots,\label{eq:omega_series}
\end{eqnarray}
where we have defined
\begin{eqnarray}
F_n(v) \equiv \frac{d^{n}f/d\phi^{n}(v)}{\left(1+\frac32\left[f'(v)/M_{Pl}\right]^2\right)^{n/2}},
\end{eqnarray}
where $n$ is the integer number and $f^{(n)}(v) \equiv d^nf/d\phi^n(v)$ is the $n$ th derivative of $f(\phi)$ with respect to $\phi$ at vev.
Since $\sigma/M_{Pl}\ll 1$ during the coherent oscillation phase after inflation, it is natural to assume the series~(\ref{eq:omega_series}) is convergent.

\begin{figure}
\includegraphics{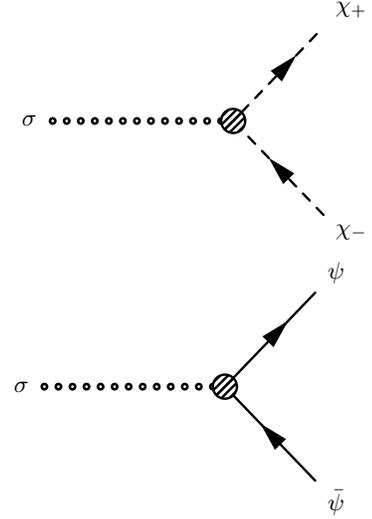}
\caption{Gravitational inflaton decay at tree level.}
\label{fig:tree}
\end{figure}

The interaction Lagrangians to the lowest order in $\sigma$ are given by expanding Eqs.~(\ref{eq:lag_scalar}) and (\ref{eq:lag_fermion}):
\begin{eqnarray}
\hat{\mathcal{L}}_{\sigma\chi\chi}&=&
\sqrt{-\hat{g}}\frac{F_1(v)}{2M_{Pl}^2}\left(\hat{g}^{\mu\nu}\hat{\chi}(D_{\mu}\hat{\chi})^* \partial_{\nu}\hat{\sigma}\right. \nonumber\\
&&\hspace{1cm}\left.+ \hat{g}^{\mu\nu}(\partial_{\mu}\hat{\sigma})\hat{\chi}^* D_{\nu}\hat{\chi} + 2m_{\chi}^2\hat{\sigma}\hat{\chi}^*\hat{\chi}\right)\nonumber\\
&&\hspace{-1cm}=
\sqrt{-\hat{g}}\frac{F_1(v)}{2M_{Pl}^2}\left(\hat{g}^{\mu\nu}\partial_{\mu}\hat{\sigma} \partial_{\nu}(\hat{\chi}^*\hat{\chi}) + 2m_{\chi}^2\hat{\sigma}\hat{\chi}^*\hat{\chi}\right)
\label{eq:lag_sigma_chi_chi}\\
&&\hspace{-1cm}\approx 
\sqrt{-\hat{g}}\frac{F_1(v)}{M_{Pl}^2}\hat{\sigma}\left(-\hat{g}^{\mu\nu}(D_{\mu}\hat{\chi})^* D_{\nu}\hat{\chi}+2m_{\chi}^2\hat{\chi}^*\hat{\chi}\right),\label{eq:lag_sigma_chi_chi_classical}\\
\hat{\mathcal{L}}_{\sigma\bar{\psi}\psi}&=& \sqrt{-\hat{g}} \frac{F_1(v)m_{\psi}}{2M_{Pl}^2}\hat{\sigma}\hat{\bar{\psi}}\hat{\psi}.
\label{eq:lag_sigma_psi_psi}
\end{eqnarray}
In deriving the last line of $\mathcal{\hat{L}}_{\sigma\chi\chi}$, we have used integration by parts and the classical equations of motion for $\hat{\chi}$ and $\hat{\chi}^*$ from Eq.~(\ref{eq:lag_scalar}): 
\begin{eqnarray}
&&\partial_{\nu}(\sqrt{-\hat{g}}\hat{g}^{\mu\nu}\partial_{\mu}\hat{\chi})\nonumber\\
&&\hspace{.5cm}=-\sqrt{-\hat{g}}\left(\Omega^{-2}m_{\chi}^2 +g^2\hat{A}^2 - 2ig \hat{A}_{\mu}\hat{g}^{\mu\nu}\partial_{\nu}\right)\hat{\chi},\nonumber\\
&&\partial_{\nu}(\sqrt{-\hat{g}}\hat{g}^{\mu\nu}\partial_{\mu}\hat{\chi}^*)\nonumber\\
&&\hspace{.5cm}=-\sqrt{-\hat{g}}\left(\Omega^{-2}m_{\chi}^2 +g^2\hat{A}^2 + 2ig \hat{A}_{\mu}\hat{g}^{\mu\nu}\partial_{\nu}\right)\hat{\chi}^*,\qquad
\end{eqnarray}
where $\partial_{\mu}\hat{A}^{\mu}=0$ has been used.
The classical field equations correspond to the on-mass shell condition in the Feynman diagrams. 
Therefore, the expression (\ref{eq:lag_sigma_chi_chi_classical}) can be used only for analysis at tree level (see Figs.~\ref{fig:tree}, \ref{fig:3body}).
For an analysis off the mass shell one has to use the expression (\ref{eq:lag_sigma_chi_chi}) instead (see Fig.~\ref{fig:loop}).
Note that neither 4-leg nor 5-leg interaction appears due to exact cancellation in the $\hat{\sigma}(D_{\mu}\hat{\chi})^*D^{\mu}\hat{\chi}$ term on the mass shell (see Fig.~\ref{fig:naive}). 
Note also that higher order interaction Lagrangians can be derived systematically by expanding Eqs.~(\ref{eq:lag_scalar}) and (\ref{eq:lag_fermion}) in power series of $\hat{\sigma}$.
In the following sections we compute exclusively in the Einstein frame. We shall suppress carets on variables when it does not cause any confusion.

\section{Inflaton decay by breaking conformal invariance}\label{sec:inflaton_decay}
\subsection{Induced decay at tree level}\label{subsec:tree_decay}

With the interaction Lagrangians, $\mathcal{L}_{\sigma\chi\chi}$ and $\mathcal{L}_{\sigma\bar{\psi}\psi}$, the inflaton decays into a light pair of scalers or fermions at tree level (Fig.~\ref{fig:tree}). 
The rates are given by the standard quantum field theory analysis \cite{Peskin:1995ev}:
\begin{eqnarray}
\Gamma(\sigma\to \chi_+\chi_-)&=& \frac{N_{\chi}\hat{g}_{\chi}^2}{4\pi m_{\sigma}}\left(1-\frac{4m_{\chi}^2}{m_{\sigma}^2}\right)^{1/2},\label{eq:decay_sigma_chi_chi}\\
\Gamma(\sigma\to \bar{\psi}\psi)&=& \frac{N_{\psi}\hat{g}_{\psi}^2m_{\sigma}}{8\pi}\left(1-\frac{4m_{\psi}^2}{m_{\sigma}^2}\right)^{3/2},\label{eq:decay_sigma_psi_psi}
\end{eqnarray}
where gravitationally induced coupling constants are defined as \cite{Watanabe:2006ku}
\begin{eqnarray}
\hat{g}_{\chi}&\equiv& \frac{F_1(v)(m_{\sigma}^2+2m_{\chi}^2)}{4M_{Pl}^2},\label{eq:g_chi}\\
\hat{g}_{\psi}&\equiv& \frac{F_1(v)m_{\psi}}{2M_{Pl}^2}.\label{eq:g_psi}
\end{eqnarray}
Note that $\hat{g}_{\chi}$ is evaluated on the mass shell, and the first term is due to the derivative coupling of scalars.\footnote{For the scalaron decay in Starobinsky's $\mathcal{R}^2$-inflation \cite{Starobinsky:1980te,Vilenkin:1985md,Mijic:1986iv,Kalara:1990ar,Faulkner:2006ub,Gorbunov:2010bn}, one can identify $F_1(v)/M_{Pl}=2/\sqrt{6}$ and $m_{\sigma}=\mu$ for the scalaron mass.}

The tree-level two-body decays are kinematically suppressed as seen in Eqs.~(\ref{eq:decay_sigma_chi_chi}) and (\ref{eq:decay_sigma_psi_psi}). Especially, the inflaton cannot decay into scalars or fermions if decay products are heavier than half mass of the inflaton, $m_{\sigma} < 2 m_{\chi},\,2 m_{\psi}$.
It is usual to have the global flavor symmetry broken, and masses of particles are different as in particle physics.
Thus, we split flavor quantum number into two pieces: 
$N_{\chi} =  N_{\chi_{\ell}}+N_{\chi_{h}}$ and 
$N_{\psi} = N_{\psi_{\ell}}+N_{\psi_{h}}$, where $N_{\chi_{\ell}}$ and $N_{\psi_{\ell}}$ denote the number of species lighter than half of the inflaton mass while $N_{\chi_h}$ and $N_{\psi_h}$ denote the number of heavier species. In this case one should replace $N_{\chi},\,N_{\psi}$ in Eqs.~(\ref{eq:decay_sigma_chi_chi}) and (\ref{eq:decay_sigma_psi_psi}) with $N_{\chi_{\ell}},\,N_{\psi_{\ell}}$.

\begin{figure}
\includegraphics{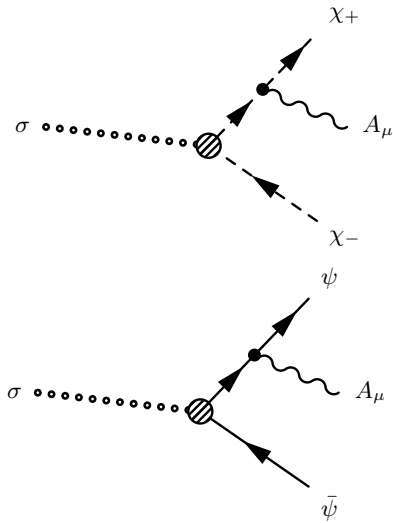}
\caption{Three-body decays at tree level.}
\label{fig:3body}
\end{figure}

At tree level the inflaton also decays into three-body (Fig.~\ref{fig:3body}) and four-body final states.
Gauge bosons are produced in the radiative decay (bremsstrahlung) processes. 
From Eq.~(\ref{eq:lag_sigma_chi_chi_classical}) one may naively expect scalar QED-like interactions shown in Fig.~\ref{fig:naive}. However, they do not appear if one evaluates the derivative couplings properly.

How fast do these processes proceed?
Three-body decays are phase-space suppressed compared to the two-body decays. For bosonic decays
\begin{eqnarray}
\Gamma(\sigma\to \chi_+\chi_-)&\simeq& \frac{N_{\chi}\hat{g}_{\chi}^2}{4\pi m_{\sigma}}
\simeq \frac{N_{\chi}[F_1(v)]^2m_{\sigma}^3}{64\pi M_{Pl}^4},\nonumber\\
\Gamma(\sigma\to \chi_+\chi_-A_{\mu})&\simeq&\frac{\alpha}{8\pi}\Gamma(\sigma\to \chi_+\chi_-),
\end{eqnarray}
where $m_{\chi}\ll m_{\sigma}$ is assumed and $\alpha\equiv g^2/(4\pi)$. 
For fermionic decays
\begin{eqnarray}
\Gamma(\sigma\to \bar{\psi}\psi) 
&\simeq& 
\frac{N_{\psi}\hat{g}_{\psi}^2m_{\sigma}}{8\pi}
= \frac{N_{\psi}[F_1(v)]^2m_{\sigma}m_{\psi}^2}{32\pi M_{Pl}^4},\nonumber\\
\Gamma(\sigma\to \bar{\psi}\psi A_{\mu}) 
&\simeq& 
\frac{\alpha}{8\pi}\Gamma(\sigma\to \bar{\psi}\psi),
\end{eqnarray}
where $m_{\psi}\ll m_{\sigma}$ is assumed.
If $g\sim \mathcal{O}(0.1)$, $\Gamma(\sigma\to 3)/\Gamma(\sigma\to 2)\simeq g^2/(32\pi^2)\sim \mathcal{O}(10^{-4}-10^{-3})$ for both $\chi$ and $\psi$.\footnote{Soft photon processes in Fig.~\ref{fig:3body} are known to be infrared log divergent. Therefore, we consider only hard photon processes in Fig.~\ref{fig:3body} by introducing a finite infrared energy cutoff for outgoing photons.} 
Although the branching ratio of three-body decay is small, pairs of decay products quickly annihilate into gauge bosons right after the gravitational inflaton decay. The annihilation process proceeds much faster than the gravitational decay if $g\sim \mathcal{O}(0.1)$.

In four-body decays, the inflaton radiates two gauge fields or another pair of charged fields in addition to a pair of charged fields. We have assumed that the gauge interaction is the only renormalizable interaction of the matter fields, $\chi$ and $\psi$; no Yukawa interaction and no self-interaction are introduced. 
If there were the Yukawa interaction in the original frame, $y\chi\bar{\psi}\psi+y\chi^*\bar{\psi}\psi$, then the four-leg interaction, $y\hat{\sigma}\hat{\chi}\hat{\bar{\psi}}\hat{\psi}+y\hat{\sigma}\hat{\chi}^*\hat{\bar{\psi}}\hat{\psi}$, would show up in the Einstein frame.
If there were the scalar self-interaction, $\lambda |\chi|^4$, the five-leg interaction, $\lambda \hat{\sigma}|\hat{\chi}|^4$, would show up. 
As is the case of three-body decays, four-body decays are phase-space suppressed compared to two-body and three-body decays.

\subsection{Induced decay via quantum trace anomaly}\label{subsec:anomaly_decay}
In this subsection, we present the main result of this paper: the rate of gravitational inflaton decay into gauge fields via gauge trace anomaly [Eq.~(\ref{eq:fs-loop})].

\begin{figure}
\includegraphics{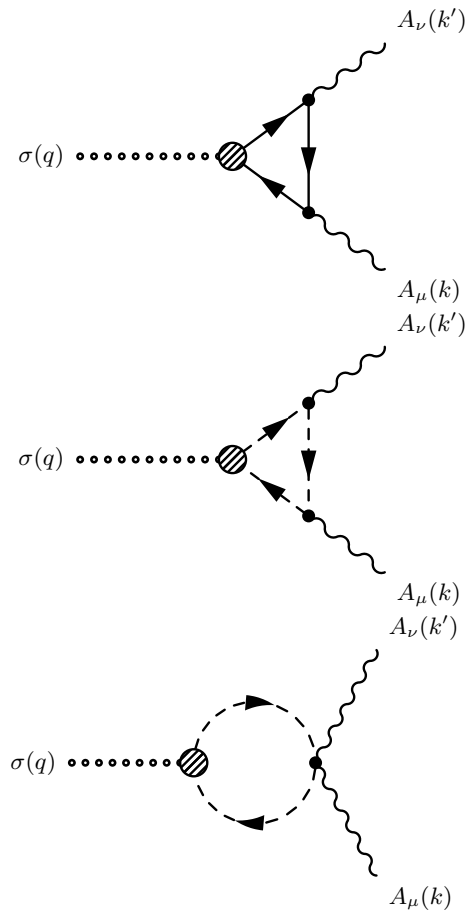}
\caption{Gravitational inflaton decay at one-loop.}
\label{fig:loop}
\end{figure}

At the classical level the inflaton can decay into only light degrees of freedom. At the quantum level, however, the inflaton can also decay into heavy degrees of freedom if they are intermediate states.
In Fig.~\ref{fig:loop} we show the two gauge boson decays of the inflaton via fermionic and bosonic one-loops.
Even though there is no induced coupling between the inflaton and massless gauge bosons, the inflaton still decays into gauge bosons through the anomaly process.

In the Einstein frame the fermionic matter field, $\psi$, couples to the inflaton by $\hat{g}_{\psi}=F_1(v)m_{\psi}/(2M_{\rm Pl}^2)$, and it also couples to gauge fields minimally. 
The fermions mediate between $\sigma$ and $A_{\mu}$ fields as a triangle loop (Fig.~\ref{fig:loop}).
We find that the decay rate via fermionic one-loop is given by
\begin{eqnarray}
\Gamma_f(\sigma\to 2A_{\mu})
=\frac{\alpha^2N^2_{\psi}[F_1(v)]^2m^3_{\sigma}}{256{\pi}^3M_{\rm Pl}^4}\left|I_f\left(\frac{m^2_{\sigma}}{m^2_{\psi}}\right)\right|^2,
\label{eq:f-loop}
\end{eqnarray}
where $\alpha\equiv g^2/(4\pi)$. It is remarkable that the intermediate particle mass dependence appears only in the function $I_f(m_{\sigma}^2/m^2_{\psi})$. A detailed derivation is given in the Appendix.
If the $\psi$ field is heavier than the inflaton, the tree-level decay [Eq.~(\ref{eq:decay_sigma_psi_psi})] is kinematically forbidden. But the decay channel to gauge bosons is still open through heavy fermions.

The decay rate via bosonic one-loop (Fig.~\ref{fig:loop}) is similarly given by
\begin{eqnarray}
\Gamma_s(\sigma\to 2A_{\mu})&&\nonumber\\
&&\hspace{-2cm}=
\frac{\alpha^2N^2_{\chi}[F_1(v)]^2m_{\sigma}^3}{1024{\pi}^3M_{\rm Pl}^4}\left(2+\frac{m_{\sigma}^2}{m_{\chi}^2}\right)^2\left|I_s\left(\frac{m^2_{\sigma}}{m^2_{\chi}}\right)\right|^2.
\label{eq:s-loop}
\end{eqnarray}
The functions, $I_f$ and $I_s$, are given in the Appendix.
In the heavy intermediate particle limit we have $I_f(m_{\sigma}^2/m_{\psi}^2 \to 0)=1/3$ and $I_s(m_{\sigma}^2/m_{\chi}^2 \to 0)=1/6$, 
and in the light intermediate particle limit
$I_f(m_{\sigma}^2/m_{\psi}^2\to\infty)=I_s(m_{\sigma}^2/m_{\chi}^2 \to\infty)= 0$ while $(2+m_{\sigma}^2/m_{\chi}^2)I_s(m_{\sigma}^2/m_{\chi}^2 )\to 2$.
Thus, only heavy intermediate particles contribute to the fermion loop [Eq.~(\ref{eq:f-loop})] while both light and heavy ones do to the scalar loop [Eq.~(\ref{eq:s-loop})].

The total decay rate for the process $\sigma\to 2A_{\mu}$ is in general given by
\begin{eqnarray}
\Gamma(\sigma\to 2A_{\mu})
=\frac{\alpha^2[F_1(v)]^2m_{\sigma}^3}{1024{\pi}^3M_{\rm Pl}^4}\left|\sum_{f=1} ^{N_{\psi}}2I_f\left(\frac{m^2_{\sigma}}{m^2_{f}}\right)\right. \nonumber\\
\left. +\sum_{s=1}^{N_{\chi}}\left(2+\frac{m_{\sigma}^2}{m_{s}^2}\right)I_s\left(\frac{m^2_{\sigma}}{m^2_{s}}\right)\right|^2.
\label{eq:fs-loop}
\end{eqnarray}
We have assumed that there is no massive gauge field at the reheating. If the energy scale of reheating is lower than that of the electroweak scale, three of four massless $B_{\mu}$ and $W^i_{\mu}$ bosons become massive $Z_{\mu}$ and $W^{\pm}_{\mu}$ bosons since the $SU(2)_L\times U(1)_Y$ gauge group is spontaneously broken down to $U(1)_{EM}$. 
To include a massive gauge boson in our model is straightforward but beyond the scope of this paper. 
The vector boson loop contribution might be as significant as scalars and fermions, but its evaluation is more involved than the present model since one has to include the Faddeev-Popov ghost field in the loop diagrams \cite{Ellis:1975ap,Vainshtein:1979}.
We will briefly discuss an effect from massive gauge bosons in the following. 

What is the physical origin of these decay processes?
If there is a dimensionful parameter, such as mass, in the Lagrangian, it breaks scale invariance explicitly. 
Conformal invariance is also broken because the scale symmetry is a subgroup of the conformal symmetry. 
The emergent interactions [Eqs.~(\ref{eq:lag_sigma_chi_chi}) and (\ref{eq:lag_sigma_psi_psi})] can be understood as a result of breaking this conformal symmetry.
 At the classical level there is no (energy) scale dependence in the dimensionless parameter, such as the gauge coupling constant. At the quantum level, however, the conformal invariance ought to be broken as the gauge coupling constant, $g$, runs with its energy scale.

The breaking of conformal invariance leads to nonzero value in the trace of the classical energy-momentum tensor \cite{Birrell:1982ix}:
\begin{eqnarray}
\hat{T}_{\rm m}^{\mu}{}_{\mu} = -\frac{\Omega}{\sqrt{-\hat{g}}}\frac{\delta \hat{S}_{\rm m}}{\delta\Omega}.
\end{eqnarray}
In deriving this expression, we use definitions of the energy-momentum tensor and functional differentiation with respect to $\Omega$:
\begin{eqnarray}
S_{\rm m} &=& {\hat{S}}_{\rm m} - \int d^4x \frac{\delta \hat{S}_{\rm m}}{\delta \hat{g}^{\mu\nu}}\delta\hat{g}^{\mu\nu} \nonumber\\
&=& {\hat{S}}_{\rm m} + \int d^4x \sqrt{-\hat{g}}\hat{T}_{\rm m}^{\mu}{}_{\mu}\delta \ln\Omega \nonumber\\
&\simeq& {\hat{S}}_{\rm m} + \int d^4x \sqrt{-\hat{g}}\hat{T}_{\rm m}^{\mu}{}_{\mu} \frac{F_1(v)\hat{\sigma}}{2M_{Pl}^2},
\label{eq:energy_momentum_tensor}
\end{eqnarray}
where we have used
\begin{eqnarray}
\hat{T}^{\rm m}_{\mu\nu}&\equiv &\frac{2}{\sqrt{-\hat{g}}}\frac{\delta \hat{S}_{\rm m}}{\delta \hat{g}^{\mu\nu}},\nonumber\\
\delta\hat{g}^{\mu\nu} &=& -2 \hat{g}^{\mu\nu}\delta\ln\Omega,
\end{eqnarray}
in the second line of Eq.~(\ref{eq:energy_momentum_tensor}), and have used Eq.~(\ref{eq:omega_series}) in the last line of Eq.~(\ref{eq:energy_momentum_tensor}).
The inflaton field, $\sigma$, thus, couples to the matter fields via the trace of the energy-momentum tensor of matter fields. 
Suppressing carets on variables again,
the leading effective interaction Lagrangian can be written as (cf. \cite{Ellis:1975ap,Vainshtein:1979})
\begin{eqnarray}
\mathcal{L}_{\rm int}=\sqrt{-g}\frac{F_1(v)\sigma}{2M_{Pl}^2} T_{\rm m}^{\mu}{}_{\mu},\hspace{4cm}\nonumber\\
T_{\rm m}^{\mu}{}_{\mu}=\sum_{s=1}^{N_{\chi}} 2\left[-(D_{\mu}\chi_s)^*D^{\mu}\chi_s+ 2U(\chi_s^*\chi_s) \right]\hspace{1.3cm}\nonumber\\
+ \sum_{f=1}^{N_{\psi}} m_f \bar{\psi}_f\psi_f
+ \sum_{v=1}^{N_V} 2m_v^2 V^{a*}_{v\mu}V^{\mu}_{va}
+ \frac{\beta_h(g)}{2g}F_{\mu\nu}F^{\mu\nu}, 
\label{eq:lag_trace}
\end{eqnarray}
where indices $f$, $s$, and $v$ run through massive fermion, scalar, and
vector field species, respectively.
We have included the contribution from massive non-Abelian gauge fields.
Here $\beta_h(g)$ is the heavy particle contribution of the Callan-Symanzik (or Gell-Mann--Low) $\beta$ function and is given by (cf. \cite{Vainshtein:1979})
\begin{eqnarray}
\label{eq:beta-function}
\beta_h(g)=-\frac{g^3}{(4\pi)^2}\sum_{heavy}\left[\left(\frac{11}{3}N-\frac13\right)N_v-\frac13 N_s -\frac23 N_f\right] \nonumber\\
+\mathcal{O}(g^5),\quad
\end{eqnarray}
where $N_s$, $N_f$, and $N_v$ are the internal (flavor) quantum number of scalar, fermion, and $SU(N)$-charged massive vector species, respectively. 
The term, $-N_v/3$, inside the bracket stands for longitudinal components of massive vector fields. 
The summation is taken over all species heavier than the inflaton, and their masses should be less than a cutoff scale of the model.
This amounts to only the one-loop correction to the effective action. The higher order loops can be ignored if the perturbation expansion converges: $g < 1$.
Since the running of $g$ is determined by effectively massless particles, $\beta_h (g)$ does not contribute to the running of $g$ until the energy scale reaches the ultraviolet region, $\mu \gtrsim m_{heavy} \gg m_{\sigma}$.

Equation~(\ref{eq:lag_trace}) takes into account only $U(1)$ charge renormalization as the leading effect to produce pairs of massless $U(1)$ gauge bosons. For a consistent one-loop analysis, one needs to include charge renormalization of $SU(N)$ gauge fields, $\beta(g_v)V_{\mu\nu}^aV^{*\mu\nu}_a/2g_v$, where $V_{\mu\nu}^a \equiv \partial_{\mu}V_{\nu}^a - \partial_{\nu}V_{\mu}^a + g_vf_{abc}V_{\mu}^bV_{\nu}^c$ and the indices $a,\,b,\,c$ take the value $+,\,-,\,3,\cdots,N^2-1$.
Renormalization of massive vector fields is beyond the scope of this paper.

In the heavy intermediate particle limit, $m_i \gg m_{\sigma}$, the decay width can be computed from the effective interaction Lagrangian [Eq.~(\ref{eq:lag_trace})].
The inflaton decays into a pair of massless gauge fields with
\begin{eqnarray}
\Gamma(\sigma\to 2A_{\mu})
\simeq \frac{\alpha^2[F_1(v)]^2m_{\sigma}^3}{1024\pi^3 M_{Pl}^4}\left|\sum_{i=s,f,v}b_i\right|^2 ,
\end{eqnarray} 
where $b_i$ are the first coefficients of the $\beta$ function [inside the bracket of Eq.~(\ref{eq:beta-function})].
The summation is taken over all species heavier than the inflaton.
With no massive vector species, $N_v = 0$, the expression agrees with Eq.~(\ref{eq:fs-loop}) in the limit of $m_i \gg m_{\sigma}$.

\section{Implication for the physics of reheating}\label{sec:implication}

What could these decay rates imply for the physics of reheating after inflation?
In general, the gauge-singlet-inflaton field has difficulty in reheating.
However, the presence of the $f(\phi)R$ term enables the singlet-inflaton to interact with matter as was shown in the previous sections. 
Therefore, the Universe reheats naturally \cite{Watanabe:2006ku}.
In order to connect the gravitational inflaton decay to the observable Universe, one actually needs to know the matter contents of the Universe after inflation.
Some of them are associated with a visible sector involving the standard model particles.
Others are associated with a hidden sector involving dark matter.
While we have not addressed these distinctions, new hints are provided by our model; the gravitational decay rate crucially depends on spins, mass spectra, and the number of degrees of freedom.

As the rates depend on the number of species in a specific way, they in principle allow us to constrain the number of heavy particles at the reheating [Eqs.~(\ref{eq:decay_sigma_chi_chi}) and (\ref{eq:decay_sigma_psi_psi}) for light particles, Eqs.~(\ref{eq:f-loop}) and (\ref{eq:s-loop}) for heavy particles]. 
Let us split the degrees of freedom into light ($m_{\ell} \ll m_{\sigma}$) and heavy ($m_{h} \gg m_{\sigma}$) species as $N_{\chi}=N_{\chi_{\ell}}+N_{\chi_h}$ and $N_{\psi}=N_{\psi_{\ell}}+N_{\psi_h}$ by assuming flavor symmetry broken into two pieces.
For scalars, in the absence of fermions, the loop process [Eq.~(\ref{eq:s-loop})] becomes faster than the tree-level process [Eq.~(\ref{eq:decay_sigma_chi_chi})] if
\begin{eqnarray}
\frac{\Gamma_s(\sigma\to 2A_{\mu})}{\Gamma(\sigma\to \chi_+\chi_-)}
\simeq
\frac{\alpha^2 (6N_{\chi_{\ell}}+N_{\chi_{h}})^2}{144\pi^2 N_{\chi_{\ell}}} >1;\label{eq:ratio_s-loop}\\
N_{\chi_h} \gtrsim 120 \frac{1}{\alpha}\left(\frac{N_{\chi_{\ell}}}{10}\right)^{1/2} \quad
{\rm if}\quad N_{\chi_h}\gg N_{\chi_{\ell}},\label{eq:n_chi_h_anomaly}\\
N_{\chi_{\ell}} \gtrsim 40 \frac{1}{\alpha^2}\quad
{\rm if} \quad N_{\chi_h}\ll N_{\chi_{\ell}}.
\label{eq:n_chi_ell_anomaly}
\end{eqnarray}
Two remarks on the condition must be made.
The first is that the channel to light scalars, including the standard model Higgs bosons, dominates over the anomaly channel if the condition [Eq.~(\ref{eq:ratio_s-loop})] is not met.
The second is that Eq.~(\ref{eq:n_chi_ell_anomaly}) holds only if the gauge interaction remains weak; it might hit the strong coupling scale with sufficiently large $N_{\chi_{\ell}}$ below or at the reheating energy scale, $T_{rh}$, since all effectively massless charged scalars contribute to the running of gauge coupling with the positive sign.
Let us assume the hierarchy of scales as $\mu\sim T_{rh} \ll m_{\sigma}$.
(If $T_{rh} \gg m_{\sigma}$, the thermal effects are important \cite{Watanabe:2006ku}.) 
Then, if $m_{\ell} \ll T_{rh}$, $N_{\chi_{\ell}}$ species contribute to $\beta$ function and the gauge coupling would become strong at $T_{rh}$ spoiling the perturbative analysis made in this paper. 
If $m_{\ell} \gg T_{rh}$, $N_{\chi_{\ell}}$ species do not contribute to $\beta$ function and the gauge coupling remains weak at $T_{rh}$.
The presence of heavy states does not matter in driving to the strong coupling scale but in enhancing the anomalous decay process.  

Similarly for fermions, in the absence of scalars $\chi_s$, the loop process [Eq.~(\ref{eq:f-loop})] becomes faster than the tree-level process [Eq.~(\ref{eq:decay_sigma_psi_psi})] if
\begin{eqnarray}
\frac{\Gamma_f(\sigma\to 2A_{\mu})}{\Gamma(\sigma\to \bar{\psi}\psi)}
\simeq
\frac{\alpha^2 m_{\sigma}^2 N_{\psi_{h}}^2}{72\pi^2 m_{\psi_{\ell}}^2 N_{\psi_{\ell}}} >1;\label{eq:ratio_f-loop}\\
N_{\psi_{h}} \gtrsim 270 \frac{m_{\psi_{\ell}}}{\alpha m_{\sigma}}\left(\frac{N_{\psi_{\ell}}}{100}\right)^{1/2},
\label{eq:n_psi_h_anomaly}
\end{eqnarray}
which can be satisfied more easily than bosonic processes since $m_{\sigma}^2 \gg m_{\psi_{\ell}}^2$. Therefore, the large number of heavy fermions may affect the reheating process significantly.

In the presence of both scalars and fermions, by comparing Eq.~(\ref{eq:fs-loop}) to Eqs.~(\ref{eq:decay_sigma_chi_chi}) and (\ref{eq:decay_sigma_psi_psi}), one finds the condition on which the anomaly induced decay is important:
\begin{eqnarray}
\frac{\Gamma(\sigma\to 2A_{\mu})}{\Gamma_{\rm tree}}\simeq
\frac{\alpha^2 }{144\pi^2}
\frac{(6N_{\chi_{\ell}}+N_{\chi_{h}}+2N_{\psi_{h}})^2}{N_{\chi_{\ell}}+2\frac{m_{\psi_{\ell}}^2}{m_{\sigma}^2}N_{\psi_{\ell}}}.
\end{eqnarray}
If there are many charged heavy species at the reheating, gauge fields can be generated efficiently via trace anomaly. Moreover if the reheating process is dominated by this channel, one can constrain the number of heavy species by the reheating temperature as in the argument of \cite{Watanabe:2006ku}; too high reheating temperature would produce unwanted relics, e.g., topological defects in the grand unified theory.

Finally, let us mention the gravitational decay in the Higgs inflation model \cite{Bezrukov:2007ep}. 
The Higgs inflaton directly couples to the matter with the gauge and Yukawa interactions.  
Since the vev of the Higgs field is small compared to the Planck scale in any sensible gauge theories, the gravitationally induced couplings are set to be small:\footnote{On the contrary, the vev of the gauge-singlet-inflaton is a free parameter, which can be $\sim M_{Pl}$.} 
$\hat{g}_{\chi}\simeq \xi v m_{\sigma}^2/(4M_{Pl}^2) \ll g m_{\chi}^2/m_{V}$ 
and $\hat{g}_{\psi}\simeq \xi v m_{\psi}/(2M_{Pl}^2) \ll y \simeq gm_{\psi}/m_{V} \lesssim \mathcal{O}(1)$, where $m_{\sigma}$ and $m_{V}$ are masses of the Higgs and massive gauge boson in the Einstein frame, respectively.
Also, $g$ and $y$ denote gauge and Yukawa coupling strengths, and are not much smaller than $\mathcal{O}(1)$. 
For simplicity, we compare perturbative decay rates for fermions:
$\Gamma_{\rm gauge}^{\rm tree}/\Gamma_{\rm grav}^{\rm tree}
\sim y_{\ell}^2 M_{Pl}^4/(\xi^2v^2m_{\sigma}m_{\psi_{\ell}}) 
\sim g^2 m_{\psi_{\ell}} M_{Pl}^4/(\xi^2v^2m_{\sigma}m_{V}^2)
\gg \mathcal{O}(1)$.\footnote{In the standard model, in fact, the decay proceeds nonperturbatively due to violation of the adiabaticity condition, and is more efficient than perturbative one \cite{GarciaBellido:2008ab,Bezrukov:2008ut}.} 
If the process is kinematically forbidden, the one-loop effect takes place instead:
$\Gamma_{\rm gauge}^{\rm one-loop}/\Gamma_{\rm grav}^{\rm one-loop}
\sim y_h^2 M_{Pl}^4/(\xi^2 v^2m_{\psi_h}^2) 
\sim g^2 M_{Pl}^4/(\xi^2 v^2m_{V}^2) 
\gg \mathcal{O}(1)$.
In both cases the ratios exhibit the hierarchy between gauge force and gravity. The main energy transfer, therefore, unlikely goes through gravitational decays unless quantum gravity becomes important.
Note that the cutoff scale is given by $\Lambda \sim M_{Pl}/(\sqrt{N}\xi)$, where $N$ is the number of species.

So far the analysis has been limited to renormalizable operators and a few of the dimension-5 operators, such as $\xi v \sigma(\partial\chi)^2$ and $\xi v \sigma F^2$. 
The smallness of $v$ made those decay channels inefficient in the standard model.
In fact, the Higgs inflation model contains up to dimension-6 operators induced by nonminimal gravitational coupling, and we have not exhausted all of them. The operators, such as  $\xi \sigma^2\chi^2$, $\xi \sigma^2\bar{\psi}\psi$, $\xi \sigma^2(\partial\chi)^2$, $\xi \sigma^2 F^2$, do not include small Higgs vev in the induced couplings, and thus might dominate over the decays from operators considered in this work. 
The gravitational pair annihilation rate of inflatons to scalars by the dimension-6 operators was calculated in \cite{Watanabe:2007tf}.

\section{Conclusion}\label{sec:conclusion}

We have studied the gravitational inflaton decays in $f(\phi)R$ gravity with the renormalizable QED-like matter sector. The model does not require any direct couplings between the inflaton and matter fields.  
We have shown that the inflaton must decay into massless gauge fields via the trace anomaly process with the rate of Eq.~(\ref{eq:fs-loop}).  
The decay channel is interesting because its amplitude is sensitive to the contribution of all the charged species in the theory including very massive ones.
Especially if the theory includes the large number of heavy charged species, they do not appear as main compositions of the primordial radiation plasma but play the role of \textit{the shadow Cabinet} enhancing the gravitational decay into gauge fields enormously.

Since the gravitational couplings are induced by breaking of conformal invariance, 
it affects all of the nonconformal fields that exist at the reheating. 
We have considered implications of the anomaly induced decay for the physics of reheating in a simple case, where the matter particles are either light ($m_i \ll m_{\sigma}$) or heavy ($m_i \gg m_{\sigma}$) compared to the inflaton. 
The fermion loop process depends only on the number of heavy species while the scalar loop process depends on that of both light and heavy species.
This new decay channel may rescue some inflation models that fail to reheat sufficiently. 
If an inflation model requires many degrees of freedom beyond the standard model, the gravitational decay via trace anomaly would contribute substantially. 
Our argument would help to constrain the nonminimal particle physics of reheating.

\begin{acknowledgments}
The author thanks Eugeny Babichev and Fedor Bezrukov for stimulating discussions and comments on the early version of the draft.
He especially thanks Eiichiro Komatsu for valuable comments and continuous encouragement to finish this work. 
This work was supported by the TRR 33 ``The Dark Universe.''
\end{acknowledgments}

\appendix
\onecolumngrid

\begin{figure}
\includegraphics{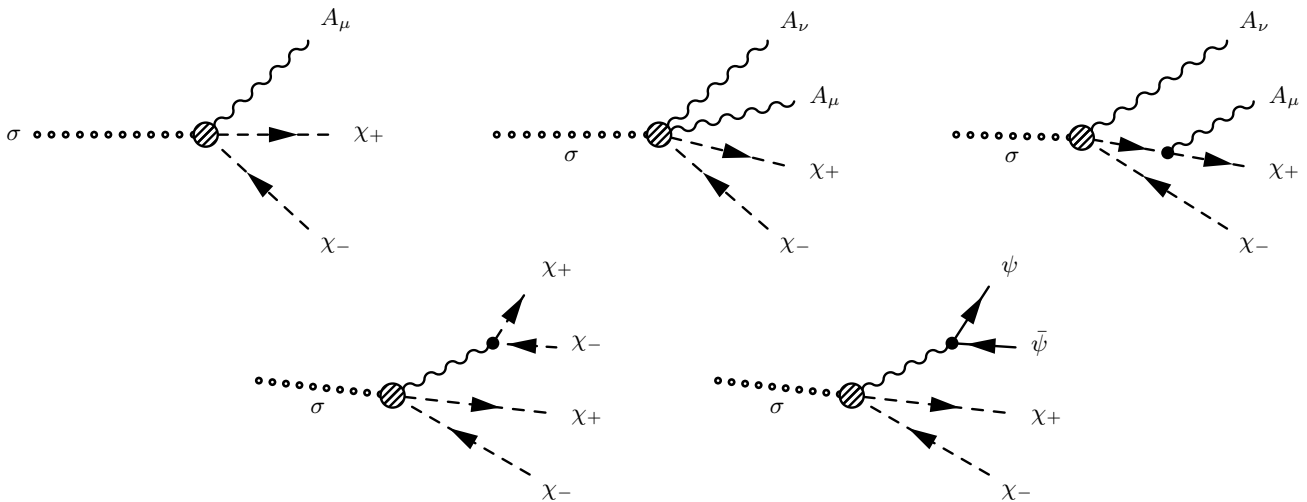}
\caption{Naive diagrams at tree-level are forbidden.}
\label{fig:naive}
\end{figure}

\section{Calculation of T-matrix amplitude for fermionic one-loop}\label{sec:loop}
We use the dimensional regularization scheme, and then the amplitude of the fermionic triangle diagram is finite.
The transition matrix ($T$-matrix) amplitude for $\sigma \to 2 A_{\mu}$ via the fermionic one-loop is 
\begin{eqnarray}
\label{eq:M-amplitude}
\langle A(k)A(k')|T|\sigma(q)\rangle &=& (2\pi)^4\delta^4(q-k-k')\mathcal{M}, \nonumber\\
\mathcal{M}
= \mathcal{M}_1 + \mathcal{M}_2 
&=& ig^2g_{f}N_{f}\epsilon^{*}_{\mu}(k)\epsilon^{*}_{\nu}(k')
\int\frac{d^4s}{(2\pi)^4}\frac{{\rm Tr} \left[(\Slash{s}+m)\gamma^{\mu}(\Slash{s}+\Slash{k}+m)\gamma^{\nu}(\Slash{s}+\Slash{q}+m)\right]}{(s^2-m^2+i\epsilon)((s+k)^2-m^2+i\epsilon)((s+q)^2-m^2+i\epsilon)}+(k\leftrightarrow k')\nonumber\\
&=& ig^2g_{f}N_{f}[\epsilon^*_{\mu}(k)\epsilon^*_{\nu}(k')I^{\mu\nu}(k,k')+\epsilon^*_{\mu}(k')\epsilon^*_{\nu}(k)I^{\mu\nu}(k',k)],\nonumber\\
I^{\mu\nu}
&\equiv& \int\frac{d^4s}{(2\pi)^4}\frac{N^{\mu\nu}}{D},
\end{eqnarray}
where we have used the Feynman slash notation for 4-momentum,
$\Slash{q}\equiv \gamma^{\mu}q_{\mu}$. 
The second term in $\mathcal{M}$ represents the triangle diagram with the opposite
charge current (or internal momentum) direction, $g_f$ is a gravitationally induced coupling constant between $\sigma$ and an intermediate fermion particle, and $m\equiv m_f$ is the mass of the intermediate particle. 

To carry out loop momentum integration we need to combine propagators. Using the Feynman parameter trick, we rewrite the denominator as
\begin{eqnarray}
\label{denominator}
\frac{1}{D}&\equiv&\frac{1}{(s^2-m^2+i\epsilon)((s+k)^2-m^2+i\epsilon)((s+q)^2-m^2+i\epsilon)}\nonumber\\
&=&2\int^1_0 dx\int^1_0 dyx \{(1-x)(s^2-m^2+i\epsilon)+xy((s+k)^2-m^2+i\epsilon)
+ x(1-y)((s+q)^2-m^2+i\epsilon) \}^{-3}\nonumber\\
&=&2\int^1_0 dx\int^1_0 dyx \{\ell^2-\Lambda^2+i\epsilon \}^{-3},
\end{eqnarray}
where
\begin{eqnarray}
\label{eq:momentum_shift}
{\ell}^{\mu}\equiv s^{\mu}+k^{\mu}x+{k'}^{\mu}x(1-y),\quad
\Lambda^2\equiv m^2-\mu^2x(1-x)(1-y).
\end{eqnarray}
Here we have used $q^{\mu}=k^{\mu}+{k'}^{\mu}$ and on-mass shell conditions: $k^2={k'}^2=0,\, q^2 \equiv\mu^2$ ($\mu^2=m_{\sigma}^2$ in the rest frame of the inflaton). The numerator in the loop integral is
\begin{eqnarray}
\label{eq:numerator}
N^{\mu\nu}&\equiv& {\rm Tr} (\Slash{s}+m)\gamma^{\mu}(\Slash{s}+\Slash{k}+m)\gamma^{\nu}(\Slash{s}+\Slash{q}+m)\nonumber\\
&=& 4m[4s^{\mu}s^{\nu}+2(s^{\mu}k^{\nu}+k^{\mu}s^{\nu}+s^{\mu}q^{\nu})+k^{\mu}q^{\nu}+q^{\mu}k^{\nu}\nonumber\\
& &\hspace{1cm}+(-2s\cdot k-k\cdot q-s^2+m^2)g^{\mu\nu}],
\end{eqnarray}
where we have used the $\gamma$-matrix algebra
\begin{eqnarray}
\label{eq:gamma-algebra}
{\rm Tr}(odd\,\#\, \gamma)&=&0,\nonumber\\
{\rm Tr}(\gamma^{\mu}\gamma^{\nu})&=&4g^{\mu\nu},\nonumber\\
{\rm Tr}(\gamma^{\mu}\gamma^{\nu}\gamma^{\alpha}\gamma^{\beta})
&=& 4(g^{\mu\nu}g^{\alpha\beta}-g^{\mu\alpha}g^{\nu\beta}+g^{\mu\beta}g^{\nu\alpha}).\quad
\end{eqnarray}
Shifting the loop momentum by Eq.~(\ref{eq:momentum_shift}), one can rewrite the numerator as
\begin{eqnarray}
N^{\mu\nu}&=& 4m[4\ell^{\mu}\ell^{\nu}+4(k^{\mu}x+{k'}^{\mu}x(1-y))(k^{\nu}x+{k'}^{\nu}x(1-y))\nonumber\\
& &-2x(3k^{\mu}k^{\nu}+k^{\mu}{k'}^{\nu})-2x(1-y)(2{k'}^{\mu}k^{\nu}+k^{\mu}{k'}^{\nu}+{k'}^{\mu}{k'}^{\nu})\nonumber\\
& &+2k^{\mu}{k}^{\nu}+k^{\mu}{k'}^{\nu}+{k'}^{\mu}k^{\nu}\nonumber\\
& &+(2x(1-y)\mu^2-\mu^2/2-\ell^2-x^2(1-y)\mu^2+m^2)g^{\mu\nu}]\nonumber\\
&=&4m[2(1-x)(1-2x)k^{\mu}k^{\nu}-2x(1-y)(1-2x(1-y)){k'}^{\mu}{k'}^{\nu}\nonumber\\
& &+(1-2x)(1-2x(1-y)){k}^{\mu}{k'}^{\nu}\nonumber\\
& &+(1-4x(1-x)(1-y)){k'}^{\mu}{k}^{\nu}\nonumber\\
& &-(1-4x(1-x)(1-y)+2x^2(1-y))\mu^2g^{\mu\nu}/2+m^2g^{\mu\nu}],
\end{eqnarray}
where the expression is valid inside the loop integral, and we have used 4-dimensional integrals in Minkowski space
\begin{eqnarray}
\int \frac{d^4\ell}{(2\pi)^4} \frac{4\ell^{\mu}\ell^{\nu}}{\{\ell^2-\Lambda^2+i\epsilon\}^3}
&=&\int \frac{d^4\ell}{(2\pi)^4} \frac{\ell^{2}g^{\mu\nu}}{\{\ell^2-\Lambda^2+i\epsilon\}^3},\nonumber\\
\int \frac{d^4\ell}{(2\pi)^4} \frac{\ell^{\mu}}{\{\ell^2-\Lambda^2+i\epsilon\}^3}&=&0.
\end{eqnarray}
Since the integrals converge, one can set $d=4$ from the start. Thus, we have evaluated the $\gamma$ matrices and loop integrations in 4-dimension.

The invariant matrix amplitude must be gauge invariant in a given order of perturbation theory. 
The Ward identity implies
\begin{eqnarray}
k_{\mu}\mathcal{M}^{\mu\nu}&=&k'_{\nu}\mathcal{M}^{\mu\nu}=0,\nonumber\\
k_{\mu}\mathcal{M}_1^{\mu\nu}&=&-k_{\mu}\mathcal{M}_2^{\mu\nu},\qquad
k'_{\nu}\mathcal{M}_1^{\mu\nu}=-k'_{\nu}\mathcal{M}_2^{\mu\nu}.
\end{eqnarray}
The invariant matrix amplitude, therefore, becomes
\begin{eqnarray}
\mathcal{M}_1 &=&ig^2g_{f}N_{f}\epsilon^*_{\mu}(k,\lambda)\epsilon^*_{\nu}(k',\lambda')I^{\mu\nu}_1,\nonumber\\
I^{\mu\nu}_1&=&\frac{-i}{(2\pi)^2m}\left(k'^{\mu}k^{\nu}-\frac{\mu^2}{2}g^{\mu\nu}\right)I\left(\frac{\mu^2}{m^2}\right),
\end{eqnarray}
where we have explicitly shown polarization of gauge fields, $\lambda$, and have used
\begin{eqnarray}
\int \frac{d^4\ell}{(2\pi)^4} \frac{1}{\{\ell^2-\Lambda^2+i\epsilon\}^3}&=&\frac{-i}{2(4\pi)^2\Lambda^2}.
\end{eqnarray}
 A function, $I_{i}(\mu^2/m_i^2)$, represents mass or energy dependence of an intermediate particle. For fermionic one-loop, 
\begin{eqnarray}
I_f(\xi)&\equiv& \int^1_0dx\int^1_0dyx\frac{1-4x(1-x)(1-y)}{1-\xi x(1-x)(1-y)}\nonumber\\
&=& \int^1_0dx\int^{1-x}_0dy\frac{1-4xy}{1-\xi xy} \nonumber\\
&=& \left\{
\begin{array}{c}
\frac{2}{\xi}\left[1+\left(1-\frac{4}{\xi}\right)\arcsin^2{\left(\frac{\sqrt{\xi}}{2}\right)}\right]\quad {\rm{if}\quad \xi \leq 4}\\
\frac{2}{\xi}\left[1-\frac14\left(1-\frac{4}{\xi}\right)\left[\ln{\left(\frac{1+\sqrt{1-4/\xi}}{1-\sqrt{1-4/\xi}}\right)} -i\pi  \right]^2  \right]\quad{\rm{if}\quad \xi > 4}
\end{array}
\right.
\end{eqnarray}
This result agrees with \cite{Steinberger:1949,Resnick:1973vg, Ellis:1975ap, Vainshtein:1979, Rizzo:1979mf}. The closed form formulas are first presented in \cite{Vainshtein:1979}.

Similarly for scalar one-loop (cf.~\cite{Ellis:1975ap,Vainshtein:1979}),
\begin{eqnarray}
I_s(\xi)&\equiv& \int^1_0dx\int^{1-x}_0dy\frac{4xy}{1-\xi xy} \nonumber\\
&=& \left\{
\begin{array}{c}
-\frac{2}{\xi}\left[1-\frac{4}{\xi}\arcsin^2{\left(\frac{\sqrt{\xi}}{2}\right)}\right] \quad {\rm{if}\quad \xi \leq 4}\\
-\frac{2}{\xi}\left[1 +\frac{1}{\xi} \left[\ln{\left(\frac{1+\sqrt{1-4/\xi}}{1-\sqrt{1-4/\xi}}\right)} -i\pi  \right]^2  \right]\quad{\rm{if}\quad \xi > 4}
\end{array}
\right.
\end{eqnarray}

The decay rate of $\sigma\to 2A_{\mu}$ can be calculated as
\begin{eqnarray}
\label{eq:decay_rate}
\Gamma(\sigma\to 2A_{\mu})&=&\frac{1}{(2\pi)^2}\frac{1}{2\mu}\frac{1}{2}\int\frac{d^3k}{2\omega_k}\int\frac{d^3k'}{2\omega_{k'}}\delta^{(4)}(q-k-k')\sum_{\lambda\lambda'}|\mathcal{M}|^2\nonumber\\
&=&\frac{\alpha^2\mu^3}{64\pi^3}\left| 
\sum_{f=1}^{N_{\psi}}\frac{\hat{g}_{f}}{m_f}I_f\left(\frac{\mu^2}{m_f^2}\right)
+\sum_{s=1}^{N_{\chi}}\frac{\hat{g}_{s}}{m_s^2}I_s\left(\frac{\mu^2}{m_s^2}\right)
\right|^2,
\end{eqnarray}
where $\alpha\equiv g^2/(4\pi)$. Note that in the heavy intermediate particle limit, we have $I_f(\xi \to 0)=1/3$ and $I_s(\xi \to 0)=1/6$, 
and in the light intermediate particle limit,
$I_f(\xi\to\infty)=I_s(\xi\to\infty)= 0$. 
One can evaluate massive vector
loop diagrams similarly; the rate should be of the
same order of magnitude as the fermion and scalar loop diagrams.

\twocolumngrid
\bibliographystyle{apsrev}
\bibliography{watanabe}

\end{document}